\documentclass[twocolumn,showpacs,amsmath,amssymb,prl]{revtex4}
\input epsf
\begin{document}
\title{Collisional Control of Ground State Polar Molecules and
Universal Dipolar Scattering}
\author{Christopher Ticknor${}^\dagger$}
\affiliation{
ARC Centre of Excellence for Quantum-Atom Optics and
Centre for Atom Optics and Ultrafast Spectroscopy,
Swinburne University of Technology, Hawthorn,
Victoria 3122, Australia }
\date{\today}
\begin{abstract}
We explore the impact of the short range interaction on the scattering
of ground state polar molecules, and
study the transition from a weak to strong dipolar scattering over  
an experimentally reasonable range of energies and electric field values.
In the strong 
dipolar limit, the scattering scales with respect to a dimensionless quantity 
defined by mass, induced dipole moment, and collision energy.  The scaling 
has implications for all quantum mechanical dipolar scattering, and therefore
this universal dipolar scaling provides estimates of scattering 
cross sections for any dipolar system. 
\end{abstract} 
\maketitle
Collisional control in ultracold atomic physics has led to the study of 
many remarkable systems such as ultracold molecules \cite{molecules}
and the BEC-BCS crossover \cite{BECBCS}.  Recently, collisional control 
of chromium has been achieved \cite{crfr}, resulting in a clear 
demonstration of magnetic dipole-dipole interaction \cite{dicold}.
Other experiments are rapidly progressing towards the production of ground
state polar molecules (GPMs), which have large permanent electric dipole 
moments.  These interact via the dipole-dipole interaction: 
\begin{eqnarray}
 &&V_{\mu\mu}={\hat{\bf \mu}_1 \cdot\hat{\bf\mu}_2 -3 (\hat {\bf R} \cdot  
\hat{\bf \mu}_1)(\hat {\bf R} \cdot \hat {\bf \mu}_2)\over R^3},
\label{fulldidi}
\end{eqnarray}
where $\hat\mu$ is the electric dipole moment of a molecule and ${\bf R}$ is 
the intermolecular separation.  Because this interaction is long range and 
anisotropic, its incorporation into many body systems has led to exciting
predictions such as dipolar crystals in restricted dimensions 
\cite{buchler} and rotons \cite{odell,shi}.  There are also many intriguing
applications of GPMs, such as quantum computing \cite{demille,micheli}.

Many different techniques are being used to obtain GPMs \cite{cold-rev},
most notably is photo-association (PA).  The PA experiments are nearing 
production of ultracold polar  ${}^1\Sigma$ molecules in their absolute 
ground states  for a range of different heteronuclear alkali-metal systems 
\cite{kerman,big,zirbel}.  ${}^1\Sigma$ molecules are relatively simple;
only the rotational structure needs to be accounted for to study the collisions
accurately. 

To set the scene, first consider ultracold atomic systems, where collisions 
are parametrized by the scattering length, $a_{s}$.  At a Feshbach 
resonance the scattering threshold becomes degenerate with a molecular
state, leading to a divergence of $a_s$.  Thus the magnetic Feshbach resonance
allows tuning of the interaction, but this is solely a short range affair.
Furthermore the only important scattering is s-wave, unless
there is another resonant partial wave, e.g., see Ref. \cite{pwave}.
This scenario is in direct contrast to dipolar scattering as will be shown.

For weakly dipolar systems, such as chromium, $a_s$ plays
a significant role in determining the dynamics of the system.  Chromium 
experiments are now exploiting Feshbach resonances to tune $a_s$ near zero; 
so dipolar interactions are dominant, and in some cases leading to
dipolar collapse \cite{dicold}.  In contrast, for strong dipolar 
scattering the short range interactions play a minor role, if not
irrelevant.  

In anticipation of these molecular systems, we have studied the scattering  
of ${}^1\Sigma$ GPMs over a wide range of electric 
fields and collision energies to determine the influence of the short 
range interaction.  This work shows the scattering of dipoles is only 
weakly influenced by the short range interaction.
We also see the emergence of a universal scaling of dipolar 
collisions which has implications for all quantum mechanical scattering 
dipoles.

To understand Eq. (\ref{fulldidi}) more clearly, consider the interaction 
in its asymptotic form in the lowest threshold containing two GPMs.  This is 
achieved when $R$ is large, and so couplings to higher thresholds are 
negligible.  This distance is quite large, typically greater than $100a_0$ for 
GPMs, where $a_0$ is the Bohr radius.  The long range interaction is 
proportional 
to the induced dipole moment, and this requires a non zero electric field.  
For simplicity of notation we use $d$ to denote the induced dipole 
moment created by a field along the $z$ axis; it is the expectation value
of the dipole moment of the field dressed ground state.
With these assumptions, the dipole-dipole interaction has the form:
${d^2\over R^3} \left(1-3\cos^2(\theta)\right)$,
where $\theta$ is the angle between ${\bf R}$ and $\hat z$. This
illustrates the  anisotropy clearly and shows that the strength of
the interaction relies directly on the induced dipole moment.  The anisotropy 
and long range nature of the interaction induce significant scattering
contributions from non-zero partial waves \cite{CT-PR,CTnew}. 
Furthermore, the long range nature of the interaction alters threshold 
behavior.  The Born approximation predicts that all partial wave cross 
sections are constant as the collision energy goes to zero \cite{threshold}.

The characteristic length and energy of a dipolar system are defined
in terms of mass $m$ (or twice the reduced mass) and the induced dipole 
moment $d$, 
and they are $D=md^2$ and $E_D=d^2/D^3=1/m^3d^4$, respectively.  With the 
scattering energy, we can form 
a dimensionless quantity $\xi=E/E_D=m^3d^4E$ which parametrizes
the scattering.  At a fixed field it is the dimensionless energy of the system.
A characteristic electric field for a $^1\Sigma$ molecule, 
${\cal E}_0$, is determined  by $B/\mu$ where $B$ is the rotational constant 
of the GPMs.

To study the GPM systems we determine the total cross section $\sigma$, 
which is  
\begin{eqnarray}
&&\sigma={2\pi\over k^2}{\cal T}\nonumber \\
&&{\cal T}=\sum_{Mij}|T^{(M)}_{ij}|^2 \label{bigt}
\end{eqnarray}
where $T_{ij}^{(M)}$ is the T matrix which details the collisions leading to
a transfer from channel $i$ to $j$ for a system with azimuthal symmetry 
$M$ about the field axis \cite{taylor}.  $k$ is the wave number, $\sqrt{mE}$.  
The factor of two in Eq. (\ref{bigt}) is there for only initially
identical scatters.  In this paper we assume identical bosons unless stated 
otherwise.  We perform scattering calculations for several polar 
molecules to obtain ${\cal T}$ and $\sigma$.   These are extremely large 
computational tasks due to the large number of partial waves
and total $M$s required to converge the calculation because of the anisotropy
and long range nature of Eq. (\ref{fulldidi}).  
The details are presented in Ref. \cite{CTnew}.  Here
we have added a Lennard-Jones potential and vary the inner wall to
alter the zero field scattering length.  The minimum of this short range 
potential, $R_{min}$, is typically 10 $a_0$ and much deeper than 
$V_{\mu\mu}(R_{min})$, thus changing the character of the scattering potential.
In zero field ($d=0$) the GPMs scatter similarly to atoms and are 
parametrized by $a_s$.

An estimate of the total cross section is achieved with a semi-classical 
approach.  This approach offers scaling of $\sigma$ on the physical parameters 
of the system, such as $d$, $m$ and $E$ \cite{hydro,rydberg}.  Another 
important cross section is the quantum unitarity limit, which 
provides a maximum value for any single partial cross section.
This occurs when the $T$ matrix takes on its maximum value of 4.  
These cross sections are:
\begin{eqnarray}
&&\sigma_{SC}={d^2\sqrt{m\over E}}c_{SC}, \label{semi}
\\
&&\sigma_Q={8 \pi\over{m E}}c_Q, \label{quantum}
\end{eqnarray}
where  $c_{SC}=1.7\times10^{-13}$ and $c_Q=4.85\times10^{-15}$
are chosen so that the units of $\mu$, $m$, $E$, and $\sigma$ 
are $[D]$, $[a.m.u.]$, $[K]$, and $[cm^2]$, respectively.


\begin{figure}
\centerline{\epsfxsize=7.0cm\epsfysize=6.0cm\epsfbox{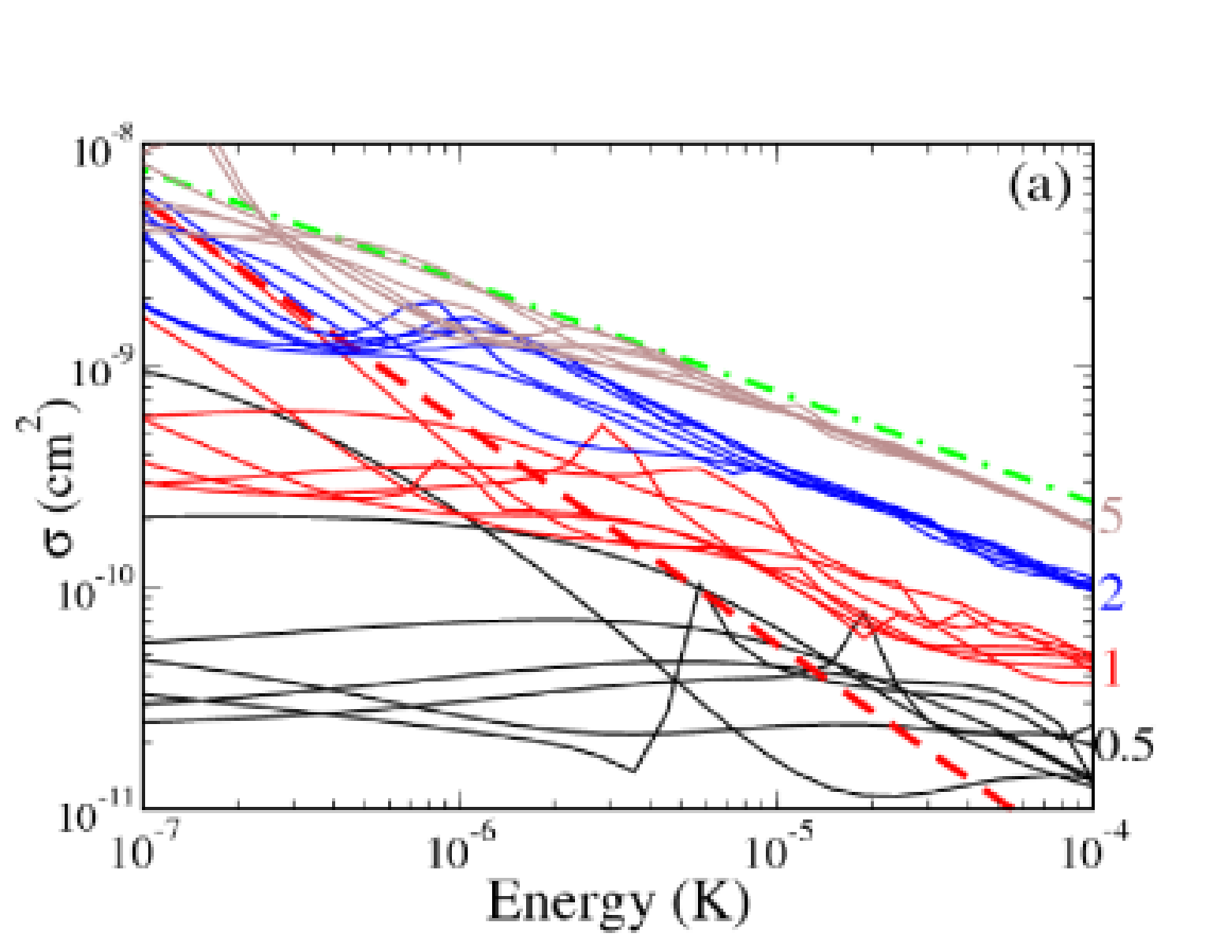}}
\centerline{\epsfxsize=7.0cm\epsfysize=6.0cm\epsfbox{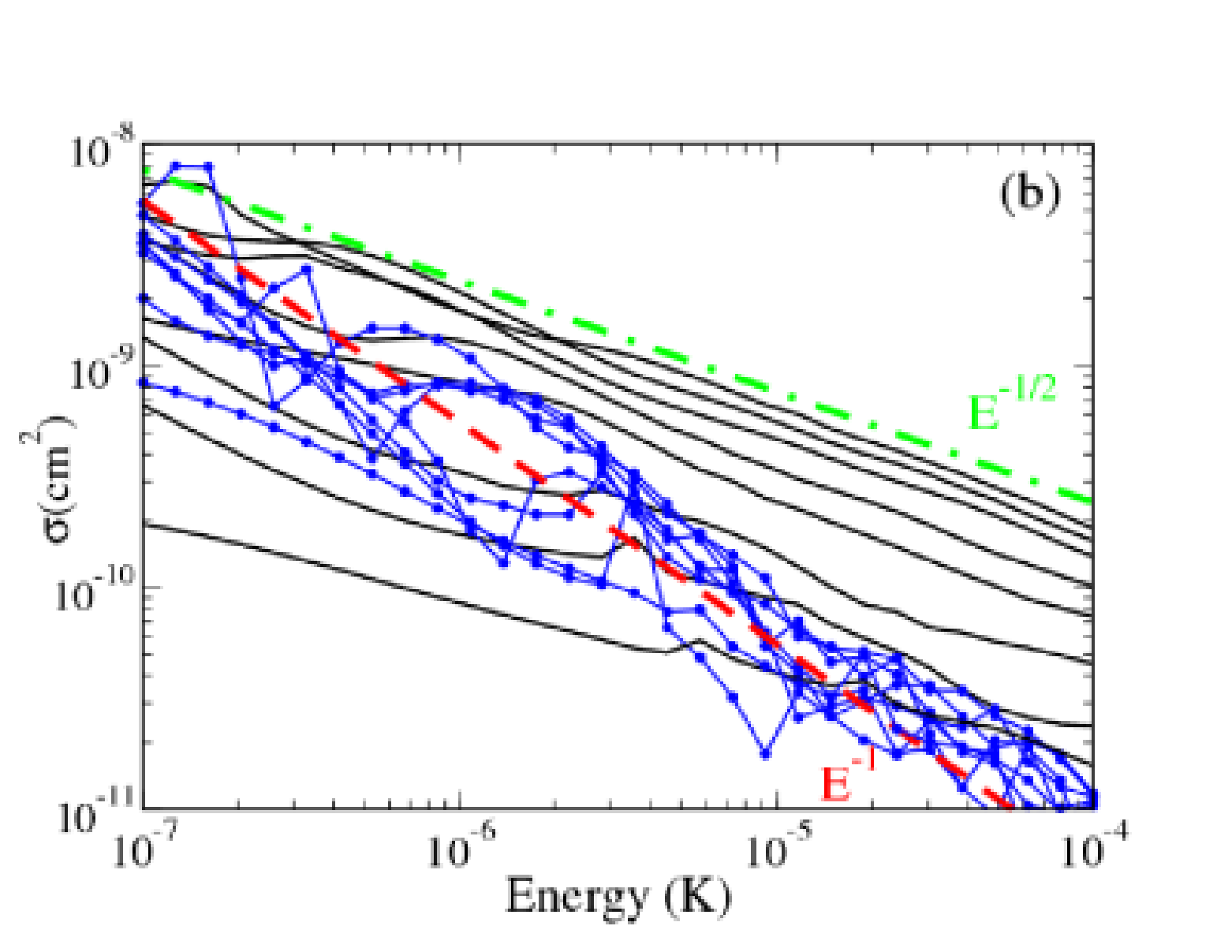}}
\caption{(Color Online) (a)  The variation of the cross section for 
many different electric fields with different short range potentials.
Different fields correspond to sets of colored curves; in ascending order the
fields are 0.5 (black), 1 (red), 2 (blue), and 5 (brown) ${\cal E}_0$.
(b) The average total cross section, $\bar\sigma$, (black) for various field 
values (in ascending order 0.5, 0.75, 1, 1.5, 2, 3, 4, and 5 ${\cal E}_0$)
and the variations, $\Delta\sigma$, about $\bar\sigma$ (blue with circles).  
In both plots the red dashed line is $\sigma_Q$ and the green dashed-dot line
is proportional to $E^{-1/2}$ as suggested by $\sigma_{SC}$.
}\label{fluxfig}
\end{figure}

To begin the analysis we illustrate the influence of the short range 
interaction as the electric field is varied.   We have calculated 
$\sigma$ of RbCs \cite{mols} as a function of energy for many different 
electric fields and with different values of $a_s$. 
Figure \ref{fluxfig} (a) shows the total RbCs cross 
section versus energy at electric fields of 
0.5 (black), 1 (red), 2 (blue), 3 (green), and 5 (brown) ${\cal E}_0$ .
For each electric field there are 7 different values of $a_s$ shown
ranging from -650 to 500 $a_0$.   The difference between these 
calculations at a given electric field is only the inner wall coefficient, 
and the calculations are identical for $R>R_{min}$.  The variation between 
calculations is due solely to the difference in short range potential, and 
therefore these calculations directly access the influence of the short 
range potential on the dipolar scattering.  The essential result of this 
figure is that at high electric field and high energy or large $\xi$
(brown) the scattering is insensitive to the short range interaction.  This 
is in contrast to low electric field and low energy or small $\xi$ (black), 
where there is great variation in the scattering due to different phases 
acquired at short range. 
 
To study this figure in detail, first we look at the weak field results
(black), small $\xi$.  The GPMs are able to access the short range interaction 
and therefore scattering is sensitive to this process.  The cross section can 
be resonantly large because $\sigma_{Q}>\sigma_{SC}$ \cite{CTnew}, 
and these resonant variations can dominate the scattering.
As the electric field is increased to 1 (red) and 2 (blue) ${\cal E}_0$, 
the cross section becomes larger and the variation in the scattering 
cross section becomes relatively small; this is most evident at high energy.
Finally, at a large electric field, 5 ${\cal E}_0$ (brown), the cross 
section is very large and there is only slight variation.
The dipole-dipole interaction has induced large numbers of 
non-zero partial waves to the scattering; we note $\sigma>>\sigma_Q$. These 
non-zero partial waves are insensitive to the short range interaction.  This 
fact constrains the resonant control of the scattering as has been seen in 
ultracold atoms with magnetic Feshbach resonances. 
It is also important to note that the cross sections do not go to zero even for
small fields, for similar reasons. 

To extent the analysis, we have averaged $\sigma$ at each field 
($\bar\sigma$) shown in 
\ref{fluxfig} (b).  In ascending order the fields for $\bar\sigma$ are 0.5, 
0.75, 1, 2, 3, 4, and 5 ${\cal E}_0$ (black).  We have also obtained
the variation of the total cross section $\Delta\sigma$,
$\Delta\sigma^2={1 \over N_a}\sum_{a=1}^{N_a} (\sigma_a-\bar\sigma)^2$.  
The variations in 
$\sigma$ are shown in Fig. \ref{fluxfig} (b) as blue full circles.  The 
largest variations occur from resonant scattering of a partial wave, and have 
a maximum contribution of $\sigma_Q$.  So it is reasonable to expect 
$\Delta\sigma\sim\sigma_Q$.  This leads to an interesting comparison, 
in the large $\xi$ limit, which is 
\begin{eqnarray}
{\Delta\sigma\over \sigma}\sim {\sigma_Q\over \sigma_{SC}}\propto 
{1\over d^2\sqrt{E}}.\label{flux}
\end{eqnarray}
This offers a more explicit statement of why for large $\xi$
there is minimal influence of the short range 
interaction.  

\begin{figure}
\centerline{\epsfxsize=8.0cm\epsfysize=7.0cm\epsfbox{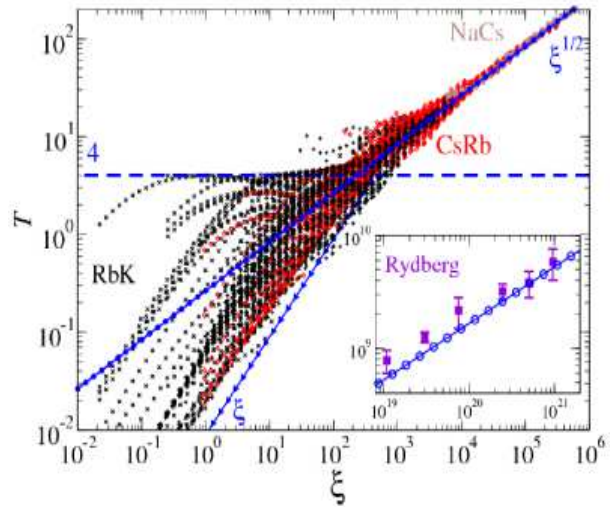}}
\caption{(Color Online) The transition of dipolar scattering to a 
universal behavior is shown by plotting $\cal T$ vs $\xi$ 
for many polar molecules.  The molecules are $^{87}$Rb$^{41}$K (black x), 
fermionic $^{87}$Rb$^{40}$K (black +), NaCs (brown $\square$), 
and RbCs (red $\diamond$).
The unitarity limit is 4, and is denoted by the dashed line.
The blue line with triangles is proportional to $\xi$.
The inset is the experimental cross section for Rydberg atoms from Ref 
\cite{rydberg}.  The blue line with open circles is the 
thermally averaged  Eq. (\ref{xifit}).
}\label{rescale}
\end{figure}
Once the short range interaction is negligible compared to the 
long range interaction (large $\xi$), it is instructive to rescale the 
multi-channel radial Schrodinger equation using the length scale $D$.
Including only the long range influences of kinetic energy and the 
dipole-dipole interaction, one finds that the only parameter left in the 
set of equations is $\xi$.  Performing the rescaling we find
\begin{equation}
\left({d^2\over dy^2}-{l(l+1)\over y^2}+\xi\right)\psi_l^{(M)}
=-\sum_{l^\prime}{C^{(M)}_{ll^\prime}\over y^3}\psi_{l^\prime}^{(M)}\label{SE}
\end{equation}
where $y=r/D$ and $\psi_l^{M}$ is a multi-channel radial wave-function.  
The coupling between partial waves, $C^{M}_{ll^\prime}$, is well known 
\cite{brink}.  Since Eq. (\ref{SE}) only depends on $\xi$, it suggests
{\it universal} scaling of dipolar scattering.  
To illustrate this behavior we have 
compiled scattering data from many different molecular systems. 
In Fig. \ref{rescale} we have plotted $\cal T$ as a function of $\xi$ 
for $^{87}$Rb$^{41}$K (black x), fermionic 
$^{87}$Rb$^{40}$K (black +), NaCs (brown squares), and RbCs (red diamonds), which
is the data in Fig. \ref{fluxfig} (a) \cite{mols}.  A line proportional to 
$\xi$ is shown as a blue line with triangles.   This line represents a cross 
section which is constant as $E\rightarrow0$ at a fixed field.
The figure shows the transition in the scattering from highly variable
at low $\xi$ to uniform at large $\xi$.  This
transition of $\cal T$ signifies the onset of universal dipolar behavior.
This will occur when the dipolar interaction is dominant  
and the scattering will be insensitive to the short range interaction.  
For this reason different molecules, even 
bosons and fermions, have the same scattering behavior.

In Eq. (\ref{quantum}), $\sigma_Q$ is defined by the $T$ matrix taking on its 
maximum value of 4.  For the same reason as $\Delta\sigma\sim\sigma_Q$, 
we find $\Delta {\cal T}\sim 4$ for all scattering.  
For small $\xi$, the scattering can access the short range and 
therefore resonant scattering is significant and so are the details of
the short range. This is seen by the great variation in the scattering
data for $\xi<200$.  
The transition to universal scattering behavior is seen as $\xi$ 
is increased above 200.  The possible values of $\cal T$ initially span a wide
range, but this span greatly decreases at high $\xi$.  This is due to 
large contributions from many non-zero partial waves; so 
typical values of $\cal T$ are much greater than the variations of any single
term.

The universal dipolar behavior is clearly seen as the scattering of the 
dipoles becomes uniform at high $\xi$.  Fitting $\cal T$ at large $\xi$, 
we find
\begin{eqnarray}
{\cal T}=0.266 \sqrt{\xi}. \label{xifit}
\end{eqnarray}
This is shown in Fig. \ref{rescale} as a blue line with full circles.
This result was obtained by fitting all RbCs and NaCs data from
$\xi=10^{4}$ to $10^{6}$, i.e., 458 scattering calculations.  
This simple equation offers
an estimate of $\cal T$ and $\sigma$ for all quantum mechanical scattering 
dipoles.  

For large $\xi$, dipolar systems obey a universal scaling, where all scattering
dipoles will behave similarly irrespective of the details of the short range. 
This implies the dipoles can be bosons, fermions, identical, or 
distinguishable, and the theory will apply.   
A striking example of this theory being applied 
is an {\it experimental} measurement of the cross section for 
resonant collisions of Rydberg atoms \cite{rydberg}.  
In this experiment two identical Rydberg atoms in the $ns$ state, 
where $n$ ($s$) is the principal quantum number (orbital angular momentum), 
are resonantly scattered into a degenerate
threshold to which it is coupled via the dipole-dipole interaction.  
For a particular electric field, the $ns+ns$ threshold becomes degenerate with 
the $np+(n-1)p$ threshold.  This system has huge dipole moments 
$d\propto n^2$, e.g., consider $n=22$, the dipole moment is about 100 $D$!
Numerically converging this calculation would be impossible with the present 
computational techniques.  But using the scaling presented here we
can obtain an accurate estimate of the total cross section.

The inset in Fig. \ref{rescale} contains the experimental cross section 
(squares) for sodium Rydberg atoms \cite{rydberg}.   
The blue line with open circles is Eq. (\ref{xifit}) with an additional 
factor of $\sqrt{\pi/2^3}$ to account for thermal averaging of
collisions in a beam \cite{metcalf}.  To calculate the values of $\xi$ for 
the experimental data we use $d=0.6{n^\star}^2$, where $n^\star$ is the 
effective quantum number, and an average collisional 
velocity of $\bar v=1.6\cdot10^{-4}$ a.u. ($T={m\over2}\bar v^2\sim0.17K$). 
The agreement of the slope is not surprising \cite{rydberg,hydro},
but agreement of the magnitude is quite good.
There has never been a means to determine the amplitude with accuracy.
The coefficient in Eq. (\ref{xifit}) allows us to accurately predict
the cross sections for {\it all} scattering dipoles.
The limit of this theory is when other physics 
emerges and alters the form of Eq. (\ref{SE}).  
For example, in many-body systems there will be screening and 
other molecules will become important before the two scattered molecules
leave the scattering volume.  

We have studied the scattering of ultracold ground
state polar molecules for experimentally accessible energies and fields.
We have illustrated the limited influence of the short-range interaction in 
the presence of dipolar interactions.  We have also rescaled many 
different scattering calculations, finding dipolar collisions are parametrized 
by $\xi=m^3d^4E$.  For our calculations $\xi$ ranges up to $10^{6}$. 
This is well into the region of universal dipolar scaling, and 
therefore we were able to determine Eq. (\ref{xifit}) with accuracy.  
This equation predicts collision cross sections for all scattering dipoles. 
Consequently we are able to estimate cross sections for quantum mechanical
dipolar systems which are far beyond our current computational capabilities.
Future directions of this work will be to study the effects of nearly 
degenerate thresholds on the scattering, such as those
presented by hyperfine structure.  

\begin{acknowledgments}
The author is grateful for support from the ARC via ACQAO and computing 
from Swinburne's Centre for Astrophysics and Supercomputing.
The author thanks C. J. Vale, A. M. Martin, and P. Hannaford for discussions 
on the manuscript.  ${}^\dagger$cticknor@swin.edu.au
\end{acknowledgments}
\bibliographystyle{amsplain}

\end{document}